# An improved, high yield method for isolating nuclei from individual zebrafish embryos for single-nucleus RNA sequencing

Clifford Rostomily[1,2,*] Heidi Lee[1,2,*] Amy Tresenrider[1,2] Riza Daza[1,2] Andrew Mullen[1,2]
Jay Shendure[1,2] David Kimelman[1,2,+] Cole Trapnell[1,2,+]

**Abstract**
      Zebrafish are an ideal system to study the effect(s) of chemical, genetic, and environmental perturbations on development due to their high fecundity and fast growth. Recently, single cell sequencing has emerged as a powerful tool to measure the effect of these perturbations at a whole embryo scale. These types of experiments rely on the ability to isolate nuclei from a large number of individually barcoded zebrafish embryos in parallel. Here we report a method for efficiently isolating high-quality nuclei from zebrafish embryos in a 96-well plate format by bead homogenization in a lysis buffer. Through head-to-head sciPlex-RNA-seq experiments, we demonstrate that this method represents a substantial improvement over enzymatic dissociation and that it is compatible with a wide range of developmental stages.

---

      Zebrafish are a useful organism for the study of gene regulation in development. Recently, multiple large scale single-cell atlases of zebrafish embryonic development have been published, allowing for the efficient annotation and analysis of new single cell datasets.[1,2] These atlases, coupled with new techniques for barcoding and multiplexing individual embryos, can be leveraged to perform high-throughput reverse genetics experiments at single-cell resolution.[2] For example, studies have leveraged these techniques to conduct time course experiments on the the effects of chemical, environmental and genetic perturbations on zebrafish development at scale.[2,3] A critical first step in multiplexed single-cell experiments is the efficient isolation of high-quality nuclei from uniquely barcoded embryos in a plate format. Previous studies used an enzymatic and mechanical dissociation approach that requires variable amounts of pipetting at elevated temperatures.[2,4] As an alternative, we developed a new multiplexed mechanical dissociation method utilizing bead homogenization in lysis buffer **(Fig. 1A, Supplementary File S1)**. Compared to enzymatic dissociation, our method is faster, less prone to experimental variability, better suited for a wider range of stages, and produces a greater number of nuclei that perform better in downstream sequencing protocols.

      In order to assess the performance of bead homogenization against enzymatic dissociation, we isolated nuclei using each method at 12 hours post-fertilization (hpf) (Experiment 1), 24 hpf, 48 hpf, and 72 hpf (Experiment 2), and 96 hpf (Experiment 3) **(Fig. 1A)**. We found that prior to sequencing, bead homogenization recovered more nuclei at each time

---

[1] University of Washington, Department of Genome Sciences, Seattle, WA, USA
[2] Seattle Hub for Synthetic Biology, Seattle, WA, USA
* These authors contributed equally to this work
+ Corresponding author

point sampled, with roughly five times more nuclei recovered at 96 hpf **(Fig. 1B)**. Enzymatic dissociation requires variable amounts of time pipetting at 37°C with longer amounts of time required in older embryos. Longer exposure to the enzymes may explain why the number of nuclei recovered does not increase after 48 hpf when using enzymatic dissociation. Nuclei recovery using bead homogenization, however, scales with developmental stage suggesting that bead homogenization may be more effective at isolating nuclei from embryos at later stages than those tested in this study **(Fig. 1B)**. To determine the effects of each method on nuclei morphology, we took images of nuclei collected at each time point. We observed that enzymatic dissociation results in small pieces of tissue that are not completely dissociated, nuclei that have tissue or extra cellular matrix still attached, and damaged nuclei that have presumably undergone too much enzymatic dissociation or experienced too much heat stress. **(Fig. 1C)**. In summary, bead homogenization recovers more nuclei per embryo and creates less debris.

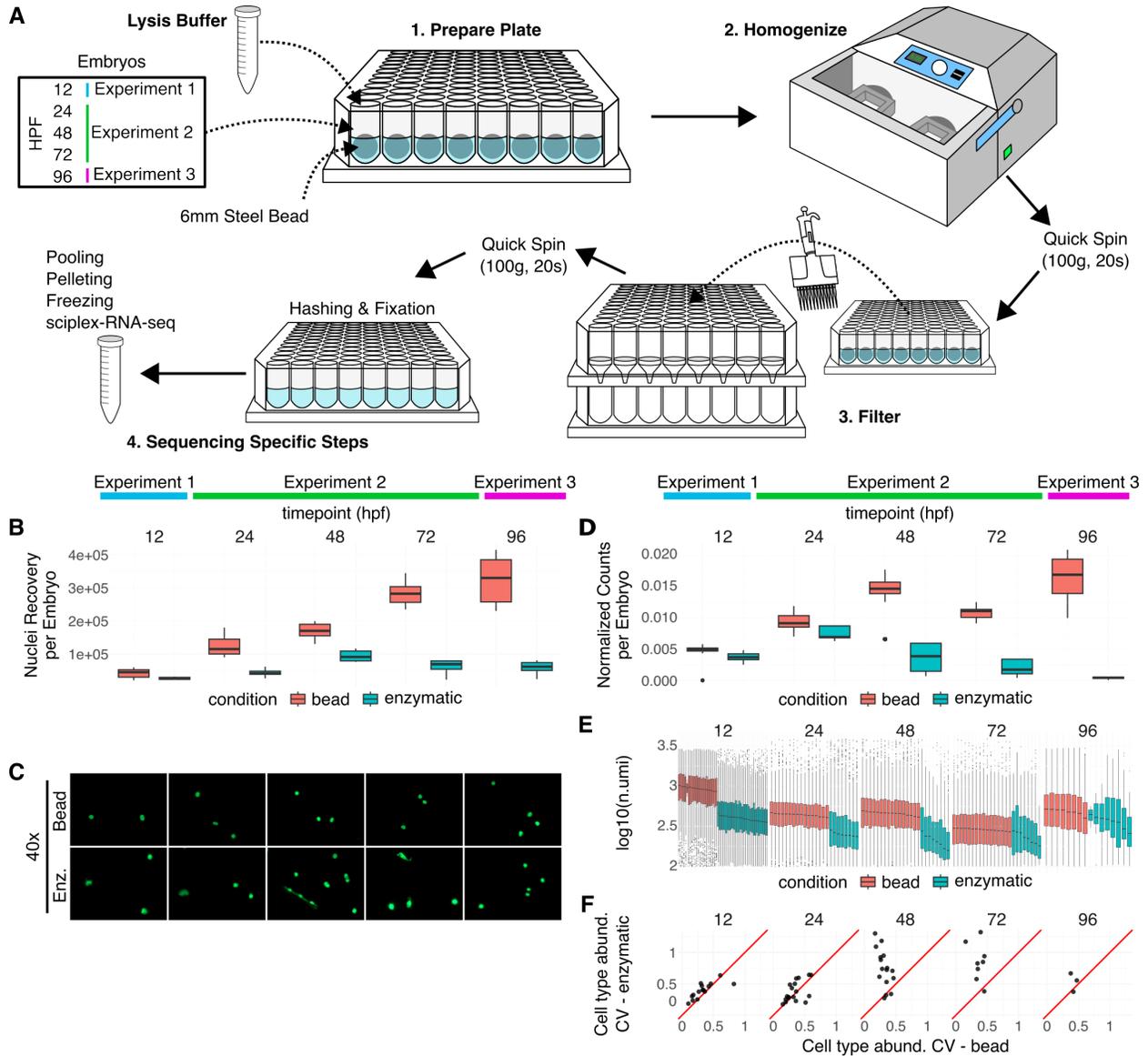

**FIG. 1. (A)** Experimental design and overview of the method. **(B)** Nuclei recovery prior to sequencing, directly after dissociation and lysis. **(C)** Images of nuclei counted in **(B)**. **(D)** Nuclei recovery after sci-plexRNAseq for bead (red) and enzymatic (blue) dissociated embryos. Counts are normalized by the number of input nuclei per embryo. **(E)** Boxplots of the UMIs recovered per embryo for bead *(red)* and enzymatic *(blue)* dissociated embryos. **(F)** Scatter plots of broad cell type coefficients of variation between bead and enzymatic dissociations.

On the nuclei isolated at each time point, we then conducted three sciPlex-RNA-seq experiments comparing our new bead homogenization method to the enzymatic method **(Fig 1A)**. Even after normalizing the number of nuclei used as input to sequencing, we detected substantially more nuclei isolated by bead homogenization, especially for older embryos **(Fig. 1D)**. From these nuclei, we also recovered more unique molecular identifiers (UMIs) per nucleus, indicating higher RNA quality **(Fig. 1E)**. Due to the multiplexing of individual embryos, we were also able to assess the embryo-to-embryo variability in cell type abundance within each experiment, and show that bead homogenization resulted in less within-experiment variability **(Fig. 1F)**. These results demonstrate that nuclei isolated by bead homogenization are more robust to the library prep, and cellular abundance is more consistent from embryo-to-embryo.

While this protocol serves as an outline for nuclear isolation by bead homogenization for sciPlex-RNA-seq in zebrafish, we anticipate that these methods can be readily adapted to additional embryo species (or embryo models), as well as to accommodate a number of other applications. For example, in preliminary tests we isolated nuclei using other lysis buffers. We have not tested nuclei isolated by bead homogenization with microfluidic-based single-cell RNAseq platforms, and additional centrifugation or filtration may be required because of the fluidics sensitivity to debris in the sample. We also suspect that this protocol will be broadly applicable to embryos from other organisms, and there is precedent for bead homogenization's use in mammals in a non-multiplexed format.[7] To facilitate optimization for new applications, we provide a troubleshooting guide of common problems we encountered and how to solve them **(Supplementary File S2)**. One barrier to the adoption of bead homogenization may be the cost of the homogenizer (~$10,000). However, in proof of concept testing, we were able to use the less expensive QIAGEN TissueLyser LT Bead Mill to dissociate embryos from tubes. This resulted in similar nuclei recovery and successful sequencing (data not shown), but it is not as amenable to multiplexed applications. Overall, the results of our experiments demonstrate that bead homogenization produces a greater number of higher quality nuclei per embryo, a greater number of UMIs per nuclei, less variability between replicates, improving the efficiency, quality, and reproducibility of multiplexed single nuclei sequencing experiments on zebrafish embryos.

## Authors Contributions

**Clifford Rostomily:** Conceptualization, Methodology, Validation, Formal analysis, Investigation, Resources, Writing - Original Draft, Writing - Review & Editing, Visualization; **Heidi Lee:** Conceptualization, Methodology, Validation, Formal analysis, Investigation, Resources, Writing - Original Draft, Writing - Review & Editing, Visualization; **Amy Tresenrider:** Methodology, Validation, Investigation, Writing - Review & Editing; **Riza Daza:** Conceptualization; **Andrew Mullen:** Investigation, Writing - Review & Editing; **Jay Shendure:** Supervision, Funding acquisition; **David Kimelman:** Resources, Writing - Review & Editing, Supervision; **Cole Trapnell:** Writing - Review & Editing, Supervision, Funding acquisition

**Acknowledgements**
We thank Dr. Lauren Saunders, Dr. Sanjay Srivatsan, and Beth Martin for helpful discussions as we were developing this method.

**Disclosures**
C.T. is a scientific advisory board member, consultant and/or co-founder of Algen Biotechnologies, Altius Therapeutics and Scale Biosciences. J.S. is a scientific advisory board member, consultant and/or co-founder of Cajal Neuroscience, Guardant Health, Maze Therapeutics, Camp4 Therapeutics, Phase Genomics, Adaptive Biotechnologies, Scale Biosciences, Sixth Street Capital, Prime Medicine, Somite Therapeutics and Pacific Biosciences. One or more embodiments of one or more patents and patent applications filed by the University of Washington may encompass methods, reagents, and the data disclosed in this manuscript. Inventors on these patents include C.T.

**Funding Statement**
This work was supported by the National Institutes of Health (RM1HG010461 to C.T., R01HG012761 to C.T. and D.K., and R01HG010632 to C.T. and J.S.) the Paul G. Allen Frontiers Group (Allen Discovery Center for Cell Lineage Tracing to C.T. and J.S.), and the Seattle Hub for Synthetic Biology, a collaboration between the Allen Institute, the Chan Zuckerberg Initiative (award number CZIF2023-008738), and the University of Washington.

**Supplementary Material (see below)**
Supplementary File S1 - Protocol
Supplementary File S2 - Troubleshooting guide
Supplementary File S3 - Hash Barcode sequences

**References**
1. Sur A, Wang Y, Capar P, et al. Single-cell analysis of shared signatures and transcriptional diversity during zebrafish development. Dev Cell 2023;58(24):3028-3047.e12; doi: 10.1016/j.devcel.2023.11.001.
2. Saunders LM, Srivatsan SR, Duran M, et al. Embryo-scale reverse genetics at single-cell resolution. Nature 2023;623(7988):782–791; doi: 10.1038/s41586-023-06720-2.
3. Dorrity MW, Saunders LM, Duran M, et al. Proteostasis Governs Differential Temperature Sensitivity across Embryonic Cell Types. 2022;2022.08.04.502669; doi: 10.1101/2022.08.04.502669.
4. VijayKumar S, Borja M, Neff N, et al. Maximizing single cell dissociation protocol for individual zebrafish embryo. Methods X 2024;13; doi: 10.1016/j.mex.2024.102958.
5. Del Priore I, Ma S, Strecker J, et al. Protocol for single-cell ATAC sequencing using combinatorial indexing in mouse lung adenocarcinoma. STAR Protoc 2021;2(2):100583; doi: 10.1016/j.xpro.2021.100583.
6. Kaya-Okur HS, Wu SJ, Codomo CA, et al. CUT&Tag for efficient epigenomic profiling of small samples and single cells. Nat Commun 2019;10(1):1930; doi: 10.1038/s41467-019-09982-5.
7. Nadelmann ER, Gorham JM, Reichart D, et al. Isolation of Nuclei from Mammalian Cells and Tissues for Single-Nucleus Molecular Profiling. Curr Protoc 2021;1(5):e132; doi: 10.1002/cpz1.132.

**Supplementary File S1 - Protocol:**

# Bead Beating Zebrafish Dissociation and Hashing with BS3/Methanol

## Materials

### Reagents

- Sodium phosphate dibasic ($Na_2HPO_4$) (Millipore Sigma, cat. no. S3264-250G)
- Sodium phosphate monobasic monohydrate ($NaH_2PO_4$-$H_2O$) (Millipore Sigma, cat. no. 71507-250G)
- Potassium phosphate monobasic ($KH_2PO_4$) (Millipore Sigma, cat. no. 60218-100G)
- Sodium chloride (NaCl) (Millipore Sigma, cat. no. S3014-500G)
- Potassium chloride (KCl) (Millipore Sigma, cat. no. P9541-500G)
- Magnesium chloride solution ($MgCl_2$), 2 M (Millipore Sigma, cat. no. 68475-100ML-F)
- Sucrose (Fisher Scientific, cat. no. S5-3)
- IGEPAL CA-630 (Millipore Sigma, cat. no. I8896-50ML)
- Diethyl pyrocarbonate (DEPC) (Millipore Sigma, cat. no. D5758-5ML)
- 10X Dulbecco's PBS (Thermo Fisher, cat. no. 14200075)
- Triton X-100 (Thermo Fisher, cat. no. A16046.AP)
- Methanol (MeOH) (Millipore Sigma, cat. no. 34860-2L-R)
- Bis(sulfosuccinimidyl)suberate (BS3) (Thermo Fisher, cat. no. 21580)
- Yoyo dye (Thermo Fisher, cat. no. Y3601)
- 100uM Hash oligos with 5' Amine Modifier C12 in the form 5'-/5AmMC12/GTCTCGTGGGCTCGGAGATGTGTATAAGAGACAG-[10bp-barcode]- BAAAAAAAAAAAAAAAAAAAAAAAAAAAAAAA-3' where B is G, C or T (IDT)
    - Refer to **Supplementary File S3** for the list of 10bp barcodes

### Equipment

- Bead Ruptor 96 (Omni International, SKU 27-0001, SKU 27-0002)
- Nunc 96-Well Polypropylene DeepWell Sample Processing & Storage Plates with Shared-Wall Technology (Thermo Fisher, cat. no. 278752)
- Nunc 96-Well Filter Plates (Thermo Fisher, cat. no. 278011)
- Breezliy 6mm Precision Steel Bearing Balls G25 304 Stainless Steel Ball (Amazon)
- Omni Sealing Mats for 2ml 96 Deep Well Plates (Millipore Sigma, cat. no. 27-530)
- Mesh strainers (Amazon)
- Refrigerated swinging bucket centrifuge with adapters to hold 96 well plates, 15ml and 50ml conical tubes, and 1.5ml microfuge tubes

- DNA/RNA LoBind tubes
- Chemical fume hood
- Multichannel pipettes and tips
- Wide bore 200ul pipette tips
- Cell counter with GFP channel

## Reagent Preparation

**10x PBS hypotonic stock solution**
1. In a 500ml beaker, dissolve the following in approx. 250ml nuclease-free water:
    a. 5.45g $Na_2HPO_4$ (dibasic)
    b. 3.1g $NaH_2PO_4$-H2O
    c. 1.2g $KH_2PO_4$
    d. 1g KCl
    e. 3g NaCl
2. Once dissolved, bring final volume up to 500ml with nuclease-free water
3. Filter sterilize the solution
4. Store at 4°C

**Lysis buffer base**
1. In a 500ml beaker, dissolve the following in approx. 250ml nuclease-free water:
    a. 50ml 10X PBS hypotonic stock solution
    b. 57g sucrose
    c. 750uL 2M $MgCl_2$
2. Once dissolved, bring the final volume up to 50ml with nuclease-free water
3. Filter sterilize the solution
4. Store at 4°C

**10% IGEPAL**
1. Mix 1ml IGEPAL CA-630 in 9ml nuclease-free water
2. Store at 4°C

**10% Triton X-100**
1. Mix 1ml Triton X-100 in 9ml nuclease-free water
2. Store at 4°C

**0.3M SPBSTM**
1. In a beaker, dissolve the following in approx. 250 nuclease-free water:
    a. 50ml 10X PBS
    b. 57g sucrose
2. Once dissolved, mix in the following:
    a. 5ml 10% Triton X-100
    b. 750ul 2M $MgCl_2$
3. Bring the final volume up to 500ml

4. Filter sterilize the solution
5. Store at 4°C

**100mM BS3**
1. Mix one 50mg vial of BS3 with 873uL nuclease-free water
2. Aliquot and store at -80°C

# Protocol

## Before Starting

- Cool swinging bucket centrifuge to 4°C
- Prepare ice buckets - after addition of lysis buffer, keep everything on ice
- Clean the silicone sealing mat if necessary (see below)
- Clean beads if necessary (see below)
- Make BS3/MeOH fixative
  - For every 5ml of fixative needed, mix 75ul 100mM BS3 with 5ml MeOH
  - Place on ice
- Make lysis buffer/IGEPAL aliquots
  - For every 1ml of lysis buffer needed, mix 1ml lysis buffer base with 10uL 10% IGEPAL and 10uL DEPC
  - Only mix the lysis buffer base and 10% IGEPAL at this time
  - Do not add in DEPC yet, DEPC has a very short half life and must be added right before use
  - Calculate the volumes needed for each aliquot based on the number of embryos plus extra to account for pipetting error
    - The first aliquot requires at least 200ul of buffer per embryo
    - The second aliquot requires at least 50ul of buffer per embryo
  - Place on ice

## Cleaning beads

1. Empty the beads from the bead plate into a catch tray
2. Transfer the beads to a 50ml conical tube
3. Add in enough 10% bleach to completely submerge the beads, invert the tube a few times and let sit for ~3-5 minutes, then pour out into the strainer
4. Rinse the beads in the strainer under running deionized water for ~1 minute.
5. Empty the beads into a clean catch tray for easier transfer into a conical tube, then transfer into a 50ml conical tube.
6. Add in enough ethanol to completely submerge the beads, invert the tube a few times, then pour out into a dry strainer.
7. Transfer the beads into a clean container and let air dry (can air dry overnight).

8. Once dry, the beads can be stored in a closed container.

## Cleaning sealing mat

1. Rinse thoroughly with 10% bleach, then DI water, then spray with 70% ethanol
2. Pat dry with a Kimwipe or let air dry, the mat can be used once it is dry

## Nuclei Isolation

1. Place a single clean 6mm steel bead into each well of a deep well bead beating plate
   - Bead beating plates with beads can be prepped ahead of time and kept covered
   - Unused wells from filters and deep well plates can still be used. Be sure to mark used wells but still load a bead into the used wells for balancing
2. Transfer embryos to a small petri dish containing clean zebrafish embryo media with 10% tricaine
3. Transfer one embryo from the small petri dish directly onto a bead using 20ul embryo media with a wide bore tip, repeat for all embryos
4. In a fume hood, add the appropriate amount of DEPC to the first aliquot of the lysis buffer/IGEPAL mix. Be sure to vortex well and work fast because DEPC has a short half-life of ~7 minutes.
   - Add 10ul of DEPC for every 1ml of lysis buffer/IGEPAL mix
   - After vortexing with DEPC, the solution should look cloudy
   - ***Note: if performing assays other than RNAseq, this lysis buffer can be substituted with a suitable alternative (e.g. OMNI-ATAC for ATACseq).
   - All subsequent steps with samples containing DEPC should be performed in a fume hood
5. Add 200ul of the lysis buffer now containing IGEPAL and DEPC to each well
6. Cover the plate with silicone sealing mat
7. Place in bead homogenizing machine and bead homogenize:
   - Make sure to use another plate containing beads as a balance in the machine
   - Recommendations for time and frequency using the Omni International Bead Ruptor 96 at different timepoints:
     i. 6 hpf - 12 hpf: 3 minutes at 3.5Hz
     ii. 18 hpf - 24 hpf: 5 minutes at 4.5Hz
     iii. 24 hpf - 96 hpf: 7 minutes at 4.5Hz
   - ***Note: Be sure to test these conditions before attempting to dissociate precious samples. Different machines or different buffers may require different settings
8. Centrifuge at 100 x g for 20 seconds at 4°C to get most of the buffer to the bottom of the well

9. Pipette the buffer up and down gently a few times while avoiding creating bubbles to resuspend nuclei, then transfer to a 20 um filter placed over a deep well catch plate
10. Add the appropriate amount of DEPC to the second aliquot of lysis buffer with IGEPAL, then add 50ul to the beads
    - Add 10ul of DEPC for every 1ml of lysis buffer/IGEPAL mix
11. Transfer to same 20 um filter while avoiding creating bubbles
12. Centrifuge filter at 100 x g for 20 seconds at 4°C to get all buffer through the filter

## sciPlex-RNA-seq Hashing, Fixing, Pooling, and Freezing

***Note: The following steps are an adaptation from the sci-RNA0seq3 and sciPlex-RNA-seq3 protocols described in Martin et al., Nature Protocols 2023, and Srivatsan et al. Science 2020. For other assays, these steps should be changed to follow the protocol of the assay being performed.

13. Add 2.5ul of 100uM hash to each well, pipette up and down gently a few times with a multichannel pipette set to 200ul using wide bore tips, then sit on ice for 5 minutes
14. Add 1000ul BS3/MeOH fixative into each well, pipette up and down gently a few times with multichannel pipette set to 200ul, then sit on ice for 15 minutes
15. Pool wells together into 50ml conical tubes and add in a volume of SPBSTM equal to 1000uL per well into the 50ml conical tubes
    - A maximum of 3 columns (24 wells) can fit into one 50ml tube
16. Centrifuge 780 x g for 15 minutes at 4°C
17. Take off supernatant, being sure not to disrupt the pellet
18. Combine the pellets using 1ml SPBSTM across all tubes
    - If you want to count the nuclei before freezing, count here using yoyo dye and a cell counter
19. Centrifuge 780 x g for 15 minutes at 4°C
20. Take off supernatant and flash freeze in liquid nitrogen

## Supplementary File S2 - Troubleshooting guide:

| Problem | Description | Solution |
|---|---|---|
| Nuclei are lysing | After staining and observing nuclei with a hemocytometer, a majority of nuclei are lysed as indicated by low nuclei counts or abnormal morphology. | Reduce frequency until lysis stops, then increase time until embryo is dissociated to satisfaction |
| Incomplete dissociation | After filtration of the dissociated embryo, a number of wells have complete embryos or large chunks left over. Note: some tissue on the filter cannot be avoided, and in our tests this does not affect the proportions of cell types recovered. | Increase frequency if nuclei are healthy until nuclei counts drop or nuclei look unhealthy. Then increase time until tissue is fully dissociated. If nuclei look unhealthy with increased time, decrease frequency and increase time again. Bead material is another variable that can be changed. We have had success with optimized conditions using glass beads with the QIAGEN TissueLyser LT as well as the optimized steel bead conditions described in this protocol for the Omni Bead Ruptor 96. |
| Large amount of debris in pellet | After pooling wells and centrifuging the pellet may contain a large amount of debris in addition to the nuclei. | For our combinatorial indexing experiments this debris has not hurt results. We freeze the pellet with the debris and briefly sonicate after thawing to dissociate the pellet before counting and proceeding with library prep. However, for some applications this debris is unacceptable and we have had success reducing centrifugation force and time until only nuclei pellet and the debris does not crash out of solution. Centrifugation can be repeated to further clean nuclei of debris. |

**Supplementary File S3 - Hash Barcode sequences**

| Embryo ID | Hash Barcode Sequence | Time Point (hpf) | Dissociation Method |
|---|---|---|---|
| expt1_bead_12_P01_A3 | TTGACTTCAG | 12 | bead |
| expt1_bead_12_P01_B3 | GGCAGGTATT | 12 | bead |
| expt1_bead_12_P01_C3 | AGAGCTATAA | 12 | bead |
| expt1_bead_12_P01_D3 | CTAAGAGAAG | 12 | bead |
| expt1_bead_12_P01_E3 | ACTCAATAGG | 12 | bead |
| expt1_bead_12_P01_F3 | CTTGCGCCGC | 12 | bead |
| expt1_bead_12_P01_G3 | AATCGTAGCG | 12 | bead |
| expt1_bead_12_P01_H3 | GGTACTGCCT | 12 | bead |
| expt1_bead_12_P01_A4 | TAGAATTAAC | 12 | bead |
| expt1_bead_12_P01_B4 | GCCATTCTCC | 12 | bead |
| expt1_bead_12_P01_C4 | TGCCGGCAGA | 12 | bead |
| expt1_bead_12_P01_D4 | TTACCGAGGC | 12 | bead |
| expt1_bead_12_P01_E4 | ATCATATTAG | 12 | bead |
| expt1_bead_12_P01_F4 | TGGTCAGCCA | 12 | bead |
| expt1_bead_12_P01_G4 | ACTATGCAAT | 12 | bead |
| expt1_bead_12_P01_H4 | CGACGCGACT | 12 | bead |
| expt1_bead_12_P01_A5 | GATACGGAAC | 12 | bead |
| expt1_bead_12_P01_B5 | TTATCCGGAT | 12 | bead |
| expt1_bead_12_P01_C5 | TAGAGTAATA | 12 | bead |
| expt1_bead_12_P01_D5 | GCAGGTCCGT | 12 | bead |
| expt1_bead_12_P01_E5 | TCGGCCTTAC | 12 | bead |
| expt1_bead_12_P01_F5 | AGAACGTCTC | 12 | bead |
| expt1_bead_12_P01_G5 | CCAGTTCCAA | 12 | bead |
| expt1_bead_12_P01_H5 | GGCGTTAAGG | 12 | bead |
| expt1_enzymatic_12_P18_A9 | GTTAGATAAG | 12 | enzymatic |
| expt1_enzymatic_12_P18_B9 | CGCTCCATGA | 12 | enzymatic |
| expt1_enzymatic_12_P18_C9 | CGAACTAAGG | 12 | enzymatic |
| expt1_enzymatic_12_P18_D9 | CCGAGGTTAA | 12 | enzymatic |
| expt1_enzymatic_12_P18_E9 | TGAACCTGAA | 12 | enzymatic |
| expt1_enzymatic_12_P18_F9 | CCGAATTAAT | 12 | enzymatic |
| expt1_enzymatic_12_P18_G9 | ATGACCATAA | 12 | enzymatic |
| expt1_enzymatic_12_P18_H9 | AAGTCAAGAT | 12 | enzymatic |
| expt1_enzymatic_12_P18_A10 | CGCCGACTGA | 12 | enzymatic |
| expt1_enzymatic_12_P18_B10 | ATGCTGAAGC | 12 | enzymatic |
| expt1_enzymatic_12_P18_C10 | TACTGGCTCA | 12 | enzymatic |
| expt1_enzymatic_12_P18_D10 | TTGATCAATA | 12 | enzymatic |
| expt1_enzymatic_12_P18_E10 | CGAACGCAAT | 12 | enzymatic |
| expt1_enzymatic_12_P18_F10 | GCGCTACGGA | 12 | enzymatic |
| expt1_enzymatic_12_P18_G10 | TTCCTCCGCA | 12 | enzymatic |
| expt1_enzymatic_12_P18_H10 | AGAGAGATCT | 12 | enzymatic |
| expt1_enzymatic_12_P18_A11 | GGCTTAGGCA | 12 | enzymatic |
| expt1_enzymatic_12_P18_B11 | CCAGACCATG | 12 | enzymatic |
| expt1_enzymatic_12_P18_C11 | ATCTTAGTCC | 12 | enzymatic |
| expt1_enzymatic_12_P18_D11 | TACTCCATCA | 12 | enzymatic |
| expt1_enzymatic_12_P18_E11 | CTTCCTTATT | 12 | enzymatic |
| expt1_enzymatic_12_P18_F11 | GGCCGACCAA | 12 | enzymatic |
| expt1_enzymatic_12_P18_G11 | AGCCTGATCC | 12 | enzymatic |
| expt1_enzymatic_12_P18_H11 | GGCGACTAAC | 12 | enzymatic |
| expt1_enzymatic_12_P18_A12 | CAATATGCGT | 12 | enzymatic |
| expt1_enzymatic_12_P18_B12 | CTAGCGAGCC | 12 | enzymatic |
| expt1_enzymatic_12_P18_C12 | AACGACTTAA | 12 | enzymatic |
| expt1_enzymatic_12_P18_D12 | ACGCTTGAGT | 12 | enzymatic |
| expt1_enzymatic_12_P18_E12 | GGCTACGTCC | 12 | enzymatic |
| expt1_enzymatic_12_P18_F12 | CTACCTCATG | 12 | enzymatic |
| expt1_enzymatic_12_P18_G12 | TTCTGCCTGG | 12 | enzymatic |
| expt1_enzymatic_12_P18_H12 | CCTGCTACCG | 12 | enzymatic |
| expt2_bead_48_A01_P18 | AGTACTTCAA | 48 | bead |
| expt2_bead_48_B01_P18 | TGCTGGATAC | 48 | bead |
| expt2_bead_48_C01_P18 | GCTGCTGAAT | 48 | bead |
| expt2_bead_48_D01_P18 | ACTGCTCCTA | 48 | bead |
| expt2_bead_48_E01_P18 | AGTTCAGTCG | 48 | bead |
| expt2_bead_48_F01_P18 | CCAGCCAGGA | 48 | bead |
| expt2_bead_48_G01_P18 | GAGCTTCAAC | 48 | bead |
| expt2_bead_48_H01_P18 | GCCTAGCTCA | 48 | bead |
| expt2_bead_48_A02_P18 | CCAATGGCCT | 48 | bead |
| expt2_bead_48_B02_P18 | AGAGAGCGAC | 48 | bead |
| expt2_bead_48_C02_P18 | GAGCGCGGCA | 48 | bead |
| expt2_bead_48_D02_P18 | AGGATGGACG | 48 | bead |
| expt2_bead_48_E02_P18 | ACGCAGTAGA | 48 | bead |
| expt2_bead_48_F02_P18 | GTAATCTTCA | 48 | bead |
| expt2_bead_48_G02_P18 | TCCATGATGA | 48 | bead |
| expt2_bead_48_H02_P18 | GCGTTGATCA | 48 | bead |
| expt2_bead_72_A04_P18 | AATTGCCGCA | 72 | bead |
| expt2_bead_72_B04_P18 | GTAGTCTGAA | 72 | bead |
| expt2_bead_72_C04_P18 | GGCGGACGTT | 72 | bead |
| expt2_bead_72_D04_P18 | CCATGATTCG | 72 | bead |
| expt2_bead_72_E04_P18 | CTCCTGGTAT | 72 | bead |
| expt2_bead_72_F04_P18 | CTGCTAGGAA | 72 | bead |
| expt2_bead_72_G04_P18 | TTCAGAGACG | 72 | bead |
| expt2_bead_72_H04_P18 | GGACTGAACT | 72 | bead |
| expt2_bead_72_A05_P18 | AATATCTCAG | 72 | bead |
| expt2_bead_72_B05_P18 | GAATTCGAGC | 72 | bead |
| expt2_bead_72_C05_P18 | AGTAAGTCGA | 72 | bead |
| expt2_bead_72_D05_P18 | GCTCATGCTT | 72 | bead |
| expt2_bead_72_E05_P18 | AATTAGAAGG | 72 | bead |
| expt2_bead_72_F05_P18 | GGCGTAGCAA | 72 | bead |
| expt2_bead_72_G05_P18 | AGATTAGCGA | 72 | bead |
| expt2_bead_72_H05_P18 | TGGCAGTTCT | 72 | bead |
| expt2_bead_24_A07_P18 | ATAGCATACT | 24 | bead |
| expt2_bead_24_B07_P18 | CCAGAGGAAG | 24 | bead |
| expt2_bead_24_C07_P18 | GGACCGGTTC | 24 | bead |
| expt2_bead_24_D07_P18 | TTGATCTCCA | 24 | bead |
| expt2_bead_24_E07_P18 | TTACCTACGT | 24 | bead |
| expt2_bead_24_F07_P18 | AAGGTCTGAT | 24 | bead |
| expt2_bead_24_G07_P18 | GGCATTATGC | 24 | bead |
| expt2_bead_24_H07_P18 | CCAGGAAGCC | 24 | bead |
| expt2_bead_24_A08_P18 | GCCTGGACTA | 24 | bead |
| expt2_bead_24_B08_P18 | AGAACCGACC | 24 | bead |
| expt2_bead_24_C08_P18 | GGAATCAGTA | 24 | bead |
| expt2_bead_24_D08_P18 | ATATGGAGAG | 24 | bead |
| expt2_bead_24_E08_P18 | AGCCATACCA | 24 | bead |
| expt2_bead_24_F08_P18 | CCTGAGGATT | 24 | bead |
| expt2_bead_24_G08_P18 | AGCCGCTCTA | 24 | bead |
| expt2_bead_24_H08_P18 | AAGATACGCA | 24 | bead |
| expt2_enz_48_A01_P01 | TTCTCGCATG | 48 | enzymatic |
| expt2_enz_48_B01_P01 | TCCTACCAGT | 48 | enzymatic |
| expt2_enz_48_C01_P01 | GCGTTGGAGC | 48 | enzymatic |
| expt2_enz_48_D01_P01 | GATCTTACGC | 48 | enzymatic |
| expt2_enz_48_E01_P01 | CTGATGGTCA | 48 | enzymatic |
| expt2_enz_48_F01_P01 | CCGAGAATCC | 48 | enzymatic |
| expt2_enz_48_G01_P01 | GCCGCAACGA | 48 | enzymatic |
| expt2_enz_48_H01_P01 | TGAGTCTGGC | 48 | enzymatic |
| expt2_enz_72_A03_P01 | TTGACTTCAG | 72 | enzymatic |
| expt2_enz_72_B03_P01 | GGCAGGTATT | 72 | enzymatic |
| expt2_enz_72_C03_P01 | AGAGCTATAA | 72 | enzymatic |
| expt2_enz_72_D03_P01 | CTAAGAGAAG | 72 | enzymatic |
| expt2_enz_72_E03_P01 | ACTCAATAGG | 72 | enzymatic |
| expt2_enz_72_F03_P01 | CTTGCGCCGC | 72 | enzymatic |
| expt2_enz_72_G03_P01 | AATCGTAGCG | 72 | enzymatic |
| expt2_enz_72_H03_P01 | GGTACTGCCT | 72 | enzymatic |
| expt2_enz_24_A05_P01 | GATACGGAAC | 24 | enzymatic |
| expt2_enz_24_B05_P01 | TTATCCGGAT | 24 | enzymatic |
| expt2_enz_24_C05_P01 | TAGAGTAATA | 24 | enzymatic |
| expt2_enz_24_D05_P01 | GCAGGTCCGT | 24 | enzymatic |
| expt2_enz_24_E05_P01 | TCGGCCTTAC | 24 | enzymatic |
| expt2_enz_24_F05_P01 | AGAACGTCTC | 24 | enzymatic |
| expt2_enz_24_G05_P01 | CCAGTTCCAA | 24 | enzymatic |
| expt2_enz_24_H05_P01 | GGCGTTAAGG | 24 | enzymatic |
| expt3_bead_96_P18_A11 | GGCTTAGGCA | 96 | bead |
| expt3_bead_96_P18_B11 | CCAGACCATG | 96 | bead |
| expt3_bead_96_P18_C11 | ATCTTAGTCC | 96 | bead |
| expt3_bead_96_P18_D11 | TACTCCATCA | 96 | bead |
| expt3_bead_96_P18_E11 | CTTCCTTATT | 96 | bead |
| expt3_bead_96_P18_F11 | GGCCGACCAA | 96 | bead |
| expt3_bead_96_P18_G11 | AGCCTGATCC | 96 | bead |
| expt3_bead_96_P18_H11 | GGCGACTAAC | 96 | bead |
| expt3_enzymatic_96_P18_A12 | CAATATGCGT | 96 | enzymatic |
| expt3_enzymatic_96_P18_B12 | CTAGCGAGCC | 96 | enzymatic |
| expt3_enzymatic_96_P18_C12 | AACGACTTAA | 96 | enzymatic |
| expt3_enzymatic_96_P18_D12 | ACGCTTGAGT | 96 | enzymatic |
| expt3_enzymatic_96_P18_E12 | GGCTACGTCC | 96 | enzymatic |
| expt3_enzymatic_96_P18_F12 | CTACCTCATG | 96 | enzymatic |
| expt3_enzymatic_96_P18_G12 | TTCTGCCTGG | 96 | enzymatic |
| expt3_enzymatic_96_P18_H12 | CCTGCTACCG | 96 | enzymatic |